\documentclass[10pt,preprint,twocolumn]{aastex631}
\usepackage{amsmath}
\usepackage{subfigure}
\usepackage{hyperref}%

 \usepackage{url}
\usepackage{graphicx}
\usepackage{natbib}
\usepackage{amssymb,txfonts}
\usepackage{multirow}
\usepackage{array}
\usepackage{sidecap}
\usepackage{hyperref}
\usepackage{epstopdf}
\usepackage{footnote}
\usepackage{tabularx}
\usepackage{booktabs}
\usepackage{soul}
\newcommand{\p}[1]{{\color{red}{#1}}}

\hypersetup{
    bookmarks=true,         
    unicode=false,          
    pdftoolbar=true,        
    pdfmenubar=true,        
    pdffitwindow=false,     
    pdfstartview={FitH},    
    pdftitle={My title},    
    pdfauthor={Author},     
    pdfsubject={Subject},   
    pdfcreator={Creator},   
    pdfproducer={Producer}, 
    pdfkeywords={keyword1} {key2} {key3}, 
    pdfnewwindow=true,      
    colorlinks=true,        
    linkcolor=blue,         
    citecolor=blue,         
    filecolor=blue
    urlcolor=blue
    }

\newcommand{\speed}[1]{#1 km~s${}^{-1}$}

\newcommand{\nfig}[1]{Figure~\ref{#1}}

\shorttitle{Formation of Double-decker Filament through Successive Tether-cutting Reconnections}
\shortauthors{Shen et al.}

\begin{document}

\title{Double-decker Pair of Flux Ropes Formed by Two Successive Tether-cutting Eruptions}
\correspondingauthor{Yuandeng Shen}
\email{ydshen@ynao.ac.cn}

\author[0000-0001-9493-4418]{Yuandeng Shen}
\affiliation{Yunnan Observatories, Chinese Academy of Sciences,  Kunming, 650216, China}

\author{Dongxu Liu}
\affiliation{Yunnan Observatories, Chinese Academy of Sciences,  Kunming, 650216, China}

\author{Surui Yao}
\affiliation{Yunnan Observatories, Chinese Academy of Sciences,  Kunming, 650216, China}

\author{Chengrui Zhou}
\affiliation{Yunnan Observatories, Chinese Academy of Sciences,  Kunming, 650216, China}

\author{Zehao Tang}
\affiliation{Yunnan Observatories, Chinese Academy of Sciences,  Kunming, 650216, China}

\author{Zhining Qu}
\affiliation{College of Physics and Electronic Engineering, Sichuan Normal University, Chengdu 610068, China}

\author{Xinping Zhou}
\affiliation{College of Physics and Electronic Engineering, Sichuan Normal University, Chengdu 610068, China}

\author{Yadan Duan}
\affiliation{Yunnan Observatories, Chinese Academy of Sciences,  Kunming, 650216, China}

\author{Song Tan}
\affiliation{Yunnan Observatories, Chinese Academy of Sciences,  Kunming, 650216, China}

\author{Ahmed Ahmed Ibrahim}
\affiliation{Department of Physics and Astronomy, College  of Science, King Saud University, P.O. Box 2455, 11451 Riyadh, Saudi Arabia}

\begin{abstract}
Double-decker filaments and their eruptions have been widely observed in recent years, but their physical formation mechanism is still unclear. Using high spatiotemporal resolution, multi-wavelength observations taken by the New Vacuum Solar Telescope and the {\em Solar Dynamics Observatory}, we show the formation of a double-decker pair of flux rope system by two successive tether-cutting eruptions in a bipolar active region. Due to the combined effect of photospheric shearing and convergence motions around the active region's polarity inversion line (PIL), the arms of two overlapping inverse-S-shaped short filaments reconnected at their intersection, which created a simultaneous upward-moving magnetic flux rope (MFR) and a downward-moving post-flare-loop (PFL) system striding the PIL. Meanwhile, four bright flare ribbons appeared at the footpoints of the newly formed MFR and the PFL. As the MFR rose, two elongated flare ribbons connected by a relatively larger PFL appeared on either side of the PIL. After a few minutes, another MFR formed in the same way at the same location and then erupted in the same direction as the first one. Detailed observational results suggest that the eruption of the first MFR might experienced a short pause before its successful eruption, while the second MFR was a failed eruption. This implies that the two newly formed MFRs might reach a new equilibrium at relatively higher heights for a while, which can be regarded as a transient double-decker flux rope system. The observations can well be explained by the tether-cutting model, and we propose that two successive confined tether-cutting eruptions can naturally produce a double-decker flux rope system, especially when the background coronal magnetic field has a saddle-like distribution of magnetic decay index profile in height.
\end{abstract}

\keywords{Solar activity(1475) --- Solar flares(1496) --- Solar filaments(1495) --- Solar magnetic reconnection(1504) --- Solar filament eruptions(1981)}

\section{Introduction} \label{introduction}
Solar filaments, known as the on-disk counterparts of prominences, are cold and dense plasma clouds suspended in the hot and tenuous corona. They appear as dark elongated linear structures located on the photospheric magnetic polarity inversion lines \citep[PIL;][]{1998SoPh..182..107M, 2010SSRv..151..333M}. The size of the filaments has a wide distribution from about $10^{3}$ to $10^{6}$ km \citep{2020RAA....20..165L}. While minifilament eruptions are capable of forming small-scale solar jets at anywhere in the lower solar atmosphere \citep[e.g.,][]{2012ApJ...745..164S, 2017ApJ...851...67S, 2021RSPSA.47700217S, 2015Natur.523..437S, 2016ApJ...830...60H, 2017ApJ...835...35H, 2020ApJ...900..158Y, 2023ApJ...942...86Y, 2023ApJ...942L..22D}, large-scale filament eruptions are the primary solar source for generating large-scale interplanetary magnetic storms that affect the near-Earth space environment \citep{2011LRSP....8....1C, 2022LRSP...19....2C, 2011RAA....11..594S, 2012ApJ...750...12S,2013ApJ...773..162B, 2015ApJ...805...48B, 2017ApJ...839..128W,2023ApJ...943...62Y}. How a filament maintains its equilibrium and stability in the corona is a key but unsolved question in filament physics. In general, it is widely accepted that the heavy filament mass could be supported by magnetic dips, which provide an upward magnetic tension force to balance the downward gravity. Magnetic dips can exist in special magnetic configurations such as sheared magnetic arcades and magnetic flux ropes (MFR). Specifically, a magnetic flux rope often refers to a magnetic structure characterized by coherently twisted magnetic field lines \citep[e.g.,][]{2020RAA....20..165L} that acts as the magnetic structure of a filament observed in H$\alpha$ or other wavelength bands and with the latter located at the bottom of the former \citep{2010ApJ...715.1566C, 2022MNRAS.516L..12T}. In this sense, a filament observed by chromospheric lines such as H$\alpha$ represents only the bottom part of an MFR. For convenience, we will use the terms filament and MFR interchangeably in this paper. In observation, the twisted magnetic structure containing a filament can be identified as intriguing bright helical threads during the eruption \citep[e.g.,][]{1987SoPh..108..251K, 2011ApJ...735L..43S, 2019ApJ...873...22S}, or during the travel of hot plasma within the MFR \citep[e.g.,][]{2019ApJ...883..104S, 2022MNRAS.516L..12T, 2023MNRAS.520.3080T}. To date, solar physicists have developed various models to explain the equilibrium and structure of filaments \citep[see,][and references therein]{2001ApJ...549.1221G, 2001ApJ...553L..85K}. Although these models can successfully explain some main features of filaments, they are difficult to interpret many new observational features based on high spatiotemporal resolution observations. For example, the simultaneous existence of both horizontal and vertical mass-carrying thin threads, peculiar bubble structures and upward moving plumes in prominences \citep[e.g.,][]{2005SoPh..226..239L, 2011Natur.472..197B, 2015ApJ...814L..17S, 2018ApJ...863..192L, 2021ApJ...923L..10C}. Recently, \cite{2012ApJ...756...59L} found that some filaments may consist of two main branches at different heights, but both are along the same PIL, and the authors called such magnetic structures `double-decker' filaments. The peculiar magnetic structure of double-decker filaments is also difficult to understand within the framework of previous models. Therefore, double-decker filaments have attracted much attention from solar physicists since their discovery, and a handful of related theoretical and observational works have been performed to explore the formation, stability, and eruption mechanisms \citep[e.g.,][]{2014ApJ...792..107K, 2019ApJ...872..109A,  2018A&A...619A.100H, 2018NewA...65....7T, 2021ApJ...909...32P, 2020ApJ...900...23M, 2022ApJ...940L..12H,2022ApJ...926..143M}. While a double-decker filament is capable of causing two successive coronal mass ejections (CMEs) from the same magnetic source region, \cite{2020ApJ...901...38J} reported that three successive CMEs launched from the same active region could be caused by the successive eruption of a ``triple-decker'' configuration consisting of three flux ropes stacked vertically above the active region PIL. \cite{2016SoPh..291.2373Z} observed the repeated formation-dissipation process of three filaments at the same place in H$\alpha$ images within 4 hours, which might also hint at the formation of a ``triple-decker'' filaments. These new observations show that the magnetic structure of filaments is much more complicated than we thought.

The physical mechanism of filament formation remains an open question \citep{2020RAA....20..166C, 2019MNRAS.488.3794W}. Previous studies have suggested that filaments could originate directly beneath the photosphere or be formed in the solar atmosphere by reconfiguring the pre-existing magnetic fields \citep[see,][and references therein]{2010SSRv..151..333M}. The first scenario implies that a pre-existing flux rope below the photosphere can partially emerge through the photosphere into the solar atmosphere due to magnetic buoyancy \citep{1994SoPh..155...69R, 2001ApJ...554L.111F}. However, direct and reliable observational evidence supporting this mechanism is still lacking, except for some possible candidate observational evidence \citep[e.g.,][]{1997SoPh..174...91L, 2010ApJ...718..474L, 2008ApJ...673L.215O, 2009ApJ...697..913O}. The latter mechanism is that sheared magnetic arcades straddling a magnetic PIL are transformed into an MFR through continuous magnetic reconnections in the lower solar atmosphere due to the combined effect of photospheric activities including shearing motions parallel to the PIL and converging motions perpendicular to the PIL \citep[][]{1989ApJ...343..971V, 2001ApJ...558..872M, 2007ApJ...666.1284W}. This physical mechanism has been supported by many numerical simulations \citep[e.g.,][]{1999ApJ...518L..57A, 2006ApJ...641..577M, 2014ApJ...780..130X, 2022ApJ...934L...9L} and multi-wavelength observations \citep[e.g.,][]{2001ApJ...560..476C, 2016ApJ...816...41Y, 2021ApJ...921L..33Y, 2015ApJS..219...17Y, 2016ApJ...832...23Y}. In addition, a large, long filament or flux rope can also be formed directly by reconfiguring two pre-existing crossed filaments or two sets of opposite J-shaped coronal loops through magnetic reconnection \citep[e.g.,][]{2007SoPh..242...53S,2009ApJ...700L..83G,2010ApJ...725L..84L, 2010IAUS..264...99L,  2010ApJ...708..314A, 2014ApJ...795....4J, 2017ApJ...845...94T, 2017ApJ...840L..23X, 2018ApJ...869...78C, 2020ApJ...902....8C}. Based on the two main filament formation mechanisms, current CME models are also divided into two categories, namely, those that consider a pre-existing flux rope emerging from below the photosphere \citep[e.g.,][]{2004ApJ...609.1123F, 2005ApJ...630L..97T}, and those that generate a flux rope during the eruption via magnetic reconnection in the corona \citep[e.g.,][]{2001ApJ...552..833M, 2021NatAs...5.1126J, 1999ApJ...510..485A, 2012ApJ...760...81K}.

For the magnetic structure of double-decker filaments, \cite{2012ApJ...756...59L} proposed two possible force-free magnetic configurations: a double flux rope equilibrium and a single flux rope above a sheared arcade. Although the two possible magnetic configurations were analytically validated to be stable \citep{2014ApJ...792..107K}, their formation mechanisms still require further observational and theoretical investigations. It has been suggested that a double-decker filament can be formed by injecting a new filament into a pre-existing one from below \citep{2012ApJ...756...59L, 2014ApJ...789..133Z, 2014SoPh..289..279Z}. In observation, the injecting filament or rising fibrils/threads often merge with the pre-existing upper filament within a few hours, and the newly merged filament may become unstable when its axial flux reaches a critical value \citep{2011ApJ...734...53S, 2014SoPh..289..279Z, 2020ApJ...898L..12Z}. A double-decker filament can also be formed by separating a pre-existing single one into two branches \citep{2012ApJ...756...59L}. \cite{2003SoPh..216..173C} noted the separation of a filament into two branches, which may be due to the presence of a positive flux within a region of negative polarity and, therefore, triggered the reconnection in the lower atmosphere and the separation of the filament. Recently, \cite{2018NewA...65....7T} reported the formation of a double-decker filament by the splitting of a large filament, in which the authors found that intermittent bright bursts in the filament channel could lead to the reconfiguration of the filament's magnetic structure. \cite{2021ApJ...909...32P} suggested that the gradual magnetic flux cancellation and converging photospheric flows around the polarity PIL could be important for filament splitting \citep{2018ApJ...860...35D, 2019ApJ...875...71Z}. In addition, some observations suggested that the splitting of a filament could be due to the magnetic reconnection inside the filament body \citep{2022ApJ...929...85D, 2022ApJ...933..200Z}. \cite{2012ApJ...750...12S} reported a clear case of reconnection within a filament-carrying flux rope system, where the authors found that a strongly twisted filament experienced a failed eruption while the cavity structure hosting the filament erupted successfully and caused a CME. This suggests that the magnetic reconnection occurred within the flux rope system but above the filament \citep{2001ApJ...549.1221G}. In other cases, it was often found that the upper filament or hot flux rope channel erupted violently as a CME, but the lower branch, which usually appears as a filament in H$\alpha$ observations, remains in its stable state \citep[e.g.,][]{2002SoPh..207..111P, 2014ApJ...789...93C, 2021ApJ...923..142C}. One problem with these observations is that the newly formed double-decker filament appears to be unstable and prone to erupt. Observations have shown that the upper branch of double-decker filaments is often observed to erupt as CMEs, and the single eruption of the lower branch \citep{2015ApJ...813...60Z}, or the successive eruptions of both branches have also been occasionally detected \citep{2018ApJ...860...35D}. 

Using high spatiotemporal resolution and multi-wavelength observations taken by the New Vacuum Solar Telescope \citep[NVST;][]{2014RAA....14..705L, 2016NewA...49....8X} and the {\em Solar Dynamics Observatory} \citep[{\em SDO};][]{2012SoPh..275....3P}, we present the observations of the successive formation and subsequent eruptions of two MFRs in the same filament channel, and our observational results might provide some new clues for understanding the formation mechanism of double-decker filaments. In this paper,  the H$\alpha$ center images were taken by the NVST, whose cadence and pixel size are 45 seconds and {0\arcsec.165}, respectively. The extreme ultraviolet (EUV) data and magnetograms were taken by the Atmospheric Imaging Assembly \citep[AIA;][]{2012SoPh..275...17L} and the Helioseismic and Magnetic Imager \citep[HMI;][]{2014SoPh..289.3483H} onboard the {\em SDO}. The AIA images have a pixel size of 0\arcsec.6 and a cadence of 12 seconds, while the cadence of the HMI magnetograms is 45 seconds.

\section{Results}
\begin{figure*}[thbp]
\plotone{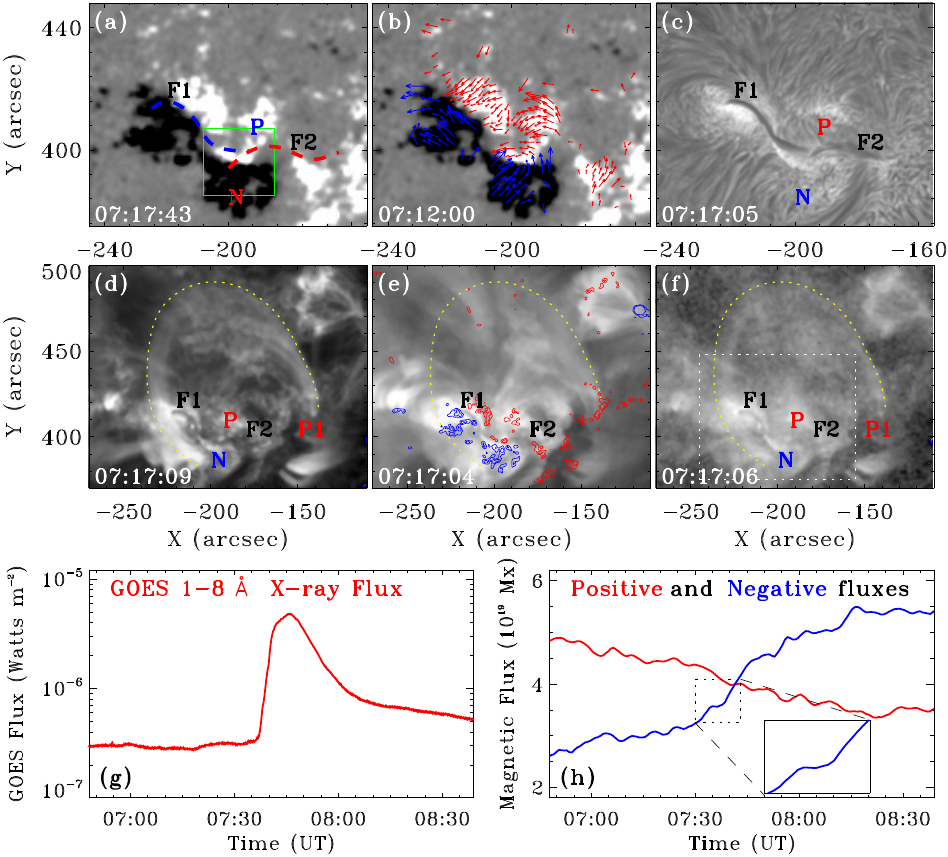}
\caption{An overview of the pre-eruption event. Panels (a) and (c) are the HMI line-of-sight magnetogram and the NVST H$\alpha$ images, respectively. The locations of the two filaments determined from the H$\alpha$ image are superimposed on the HMI line-of-sight magnetogram as blue and red curves, respectively. Panel (b) is an HMI line-of-sight magnetogram overlaid with the calculated photospheric velocity field as red and blue arrows, where the direction and length of the red and blue arrows indicate the moving direction and velocity magnitude of the positive and negative magnetic elements, respectively. Panels (d) -- (f) are AIA 171 \AA\, 193 \AA\ and 131 \AA\ images, respectively. The yellow dotted curve outlines the overlying large loop system, while the white dashed rectangle shows the field-of-view of panels (a) -- (c). The red and blue curves superimposed in panel (e) are the contours of the HMI line-of-sight magnetogram. In each panel, the letters `P' and `N' denote the central positive and negative magnetic polarities, `P1' denotes the positive magnetic polarity at the north footprint of the large loop system, and `F1' and `F2' denote the two small filaments, respectively. Panel (g) shows the {\em GOES} soft X-ray flux of the 1 -- 8 \AA\ channel. Panel (h) shows the magnetic flux within the green rectangle in panel (a), where the red and blue curves are the plots of the positive and the absolute values of the negative magnetic fluxes, respectively. The inset in panel (h) is a close-up view of the negative magnetic flux curve from 07:30 UT to 07:43 UT.
\label{fig1}}
\end{figure*}

The event occurred in NOAA active region AR12971 on 2022 March 20, and it was accompanied by a {\em GOES} C4.6 flare with start, peak, and end times at 07:30 UT, 07:45 UT, and 07:53 UT, respectively (see \nfig{fig1} (g)). The eruption source region consisted of two regions of opposite magnetic polarities separated by an inversed S-shaped PIL. As shown in \nfig{fig1}, the southeastern part of the active region was of negative magnetic polarity (black patches), while the northwestern part was of positive magnetic polarity  (white patches, see \nfig{fig1} (a)). There are two small overlapping inverse-S-shaped filaments (F1 and F2) can be observed in the pre-eruption H$\alpha$ image along the PIL (see \nfig{fig1} (c) and the red and blue dashed lines in \nfig{fig1} (a)). Before the eruption, the two opposite magnetic polarities in the HMI magnetogram showed obvious convergence motion perpendicular to the PIL and relatively weak shearing motion parallel to the PIL. Using the Differential Affine Velocity Estimator for Vector Magnetograms technique \citep[DAVE4VM;][]{2008ApJ...683.1134S} and the photospheric vector magnetic field data at 07:00:18 UT and 07:12:18 UT, the photospheric mean velocities of the positive and negative magnetic polarities were calculated to be \speed{0.25} and \speed{0.43}, respectively. The calculated velocity field is superimposed on the HMI line-of-sight magnetogram at 07:12:18 UT as red and blue arrows in \nfig{fig1} (b), where the direction and length of the arrows represent the direction and magnitude of the moving magnetic elements. The strongest photospheric motion occurred around the intersection between F1 and F2. Therefore, we further investigated the variations of the positive and negative magnetic fluxes within this region (see the green rectangle in \nfig{fig1} (a)), and the result is plotted in \nfig{fig1} (h). It can be seen that as the negative magnetic flux increased, the positive magnetic flux decreased significantly over time. In addition, by seeing the time sequence HMI line-of-sight magnetograms, the positive and negative polarities show obvious simultaneous shear and convergence motions and annihilate at the middle section of the PIL. Such a magnetic flux variation pattern suggests the magnetic flux cancellation between the positive and negative magnetic fluxes due to the combined effect of shear and convergence motions of the opposite magnetic polarities. It is noted that during the period of fast increase of the negative magnetic flux, the rate of increase slowed down or stopped for about 5 minutes from about 07:34 UT to 07:39 UT, after which the increase of the negative magnetic flux within the box region continued at a regular rate as before (see the inset in \nfig{fig1} (h)). The AIA 171 \AA, 193 \AA, 131 \AA\ images show the coronal condition of the eruption source region (see \nfig{fig1} (d)--(f)), which exhibited not only the two small filaments but also a large overlying coronal loop connecting polarities N and P1. 

\begin{figure*}[thbp]
\plotone{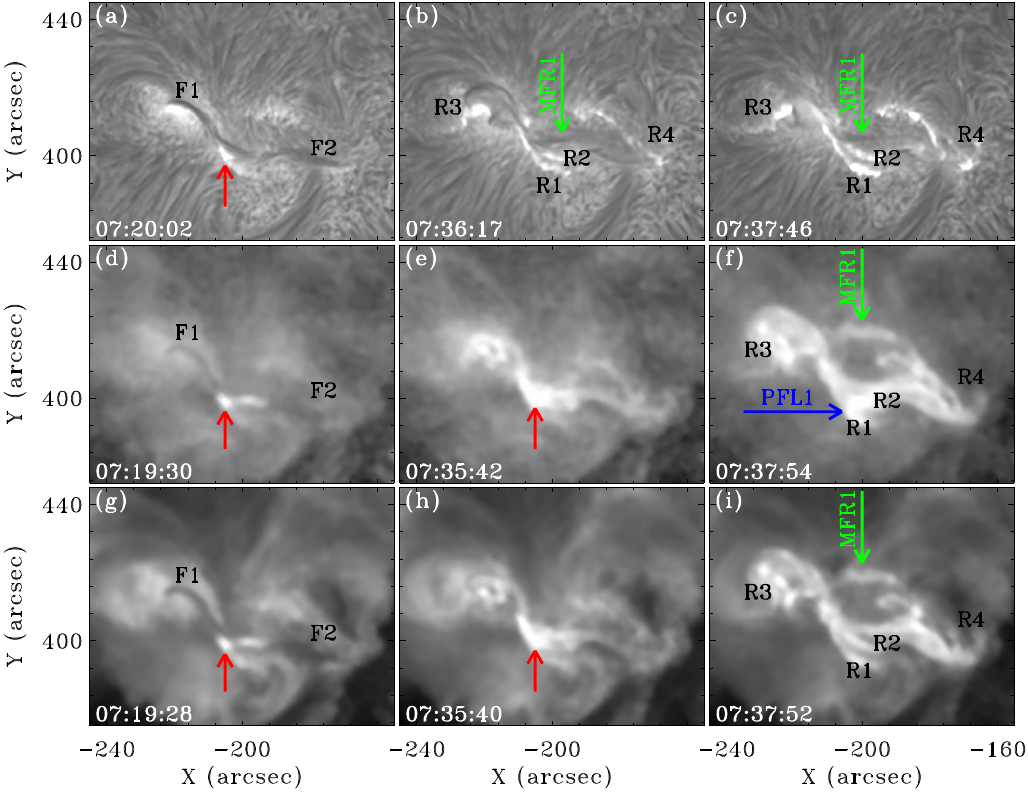}
\caption{Formation and eruption of MFR1. The top, middle, and bottom rows are the NVST H$\alpha$, the AIA 131 \AA\, and 193 \AA\ time series images, respectively. The red arrows point to the brightening around the crossed location between F1 and F2. The green arrows in panels (b), (c), (f), and (i) indicate the rising MFR1. F1 and F2 are denoted by the symbols `F1' and `F2', while the four flare ribbons are denoted by `R1' -- `R4', respectively. `PFL1' denotes the flare loop system that strides the PIL during the eruption of MFR1.
\label{fig2}}
\end{figure*}

\begin{figure*}[thbp]
\plotone{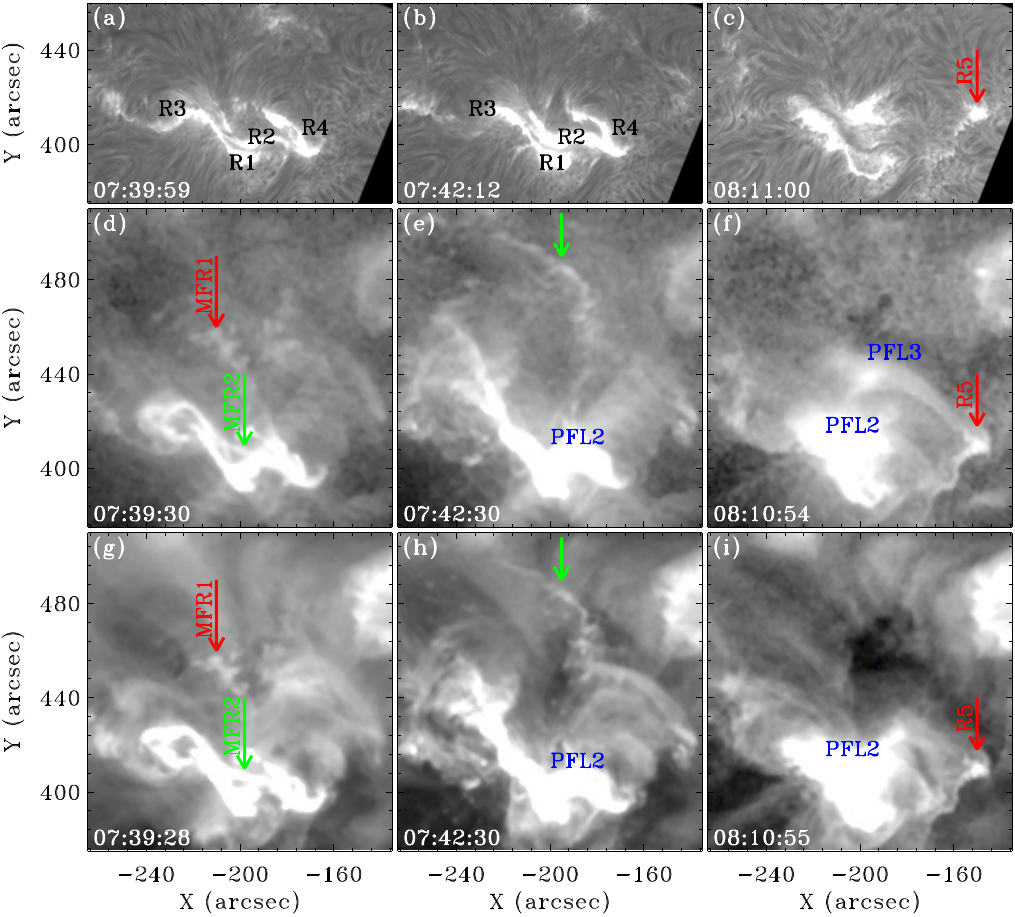}
\caption{Formation and eruption of MFR2. The top, middle, and bottom rows are the NVST H$\alpha$, the AIA 131 \AA\, and 193 \AA\ time series images, respectively. The red arrows in panels (d) and (g) point to the erupting MFR1, while the green arrows in panels (d), (e), (g), and (h) point to the erupting MFR2. The red arrows in panels (c), (f), and (i) indicate the flare ribbon R5. PFL2 indicates the flare loop associated with the eruption of MFR2, and PFL3 indicates the other group of flare loops. As in \nfig{fig2}, the four flare ribbons are also denoted by the symbols `R1' -- `R4', respectively. This figure is accompanied by an animation showing the time evolution of the event in the NVST H$\alpha$ (upper left), the HMI (lower left), the AIA 131 \AA\ (upper right) and the AIA 193 \AA\ (lower right) wavelength bands from 07:10:16 UT to 08:30:54 UT. The duration of the animation is 17 s, the field of views of the NVST H$\alpha$, AIA 131 \AA\ and 193 \AA\ images are the same as in this figure,  and the field of view of the HMI magnetogram is the same as with the NVST H$\alpha$ in this figure. Annotations in the figure are not plotted in the animation.
\label{fig3}}
\end{figure*}

The eruption details of the event are presented in \nfig{fig2} and \nfig{fig3}, using the NVST H$\alpha$ center, AIA 131 \AA\ and 193 \AA\ time sequence images (an animation is available in the online journal). In the present event, it is interesting that we observed the successive formation and subsequent eruption of two MFRs due to the magnetic reconnection between F1 and F2 at their crossing location. The formation and eruption of the first MFR is displayed in \nfig{fig2}. In the H$\alpha$ images (the top row of \nfig{fig2}), an obvious bright patch appeared at the intersection between F1 and F2 at about 07:20:02 UT before the appearance of the MFR (see the red arrow in \nfig{fig2} (a)). After that, F1 and F2 were activated and became thicker and darker. At about 07:34 UT, a dark and long filament (we will call it MFR1 hereafter) connecting R3 and R4 appeared and started to rise, and this process can be observed in the H$\alpha$ images at 07:36:17 UT and 07:37:46 UT (see the green arrows in \nfig{fig2} (b) and (c) and the online animation). In the meantime, four bright flare ribbons appeared simultaneously with the appearance of MFR1 (see the symbols for R1, R2, R3, and R4 in \nfig{fig2} (b) and (c)), two of which (R1 and R2) appeared at both sides of the PIL and the other two (R3 and R4) appeared at the far ends of F1 and F2, respectively. In the AIA 131 \AA\ images (the middle row of \nfig{fig2}), one can also see the appearance of the brightening at the intersection between F1 and F2 (see the red arrow in \nfig{fig2} (d) and (e)), the four flare ribbons, and the rising MFR1 (see the green arrow in \nfig{fig2} (f)). New signals in the AIA 131 \AA\ are that the newly formed MFR1 was bright rather than dark as it appeared in the H$\alpha$ images (see the green arrow in \nfig{fig2} (f)), and a bright flare loop system (hereafter, we call it PFL1) connecting R1 and R2 can be identified clearly (see the blue arrow in \nfig{fig2} (f)). In AIA 193 \AA\ images (bottom row of \nfig{fig2}), the brightening patch at the intersection between F1 and F2, the four flare ribbons, and the rising MFR1 can also be observed. However, one can not identify the flare loop connecting R1 and R2 as shown in AIA 131 \AA\ image at 07:37:54 UT. These observational characteristics suggest that the eruption process can be well explained by the so-called tether-cutting model \citep{2001ApJ...552..833M, 2021NatAs...5.1126J} which has been confirmed by many observational studies \citep{2013ApJ...778L..36L, 2014ApJ...797L..15C, 2016ApJ...818L..27C, 2018ApJ...869...78C, 2021ApJ...915...55L}. Here, MFR1 appeared as dark in H$\alpha$ but bright in AIA 131 \AA\ images, which might suggest that the newly formed MFR1 was composed of both cool and hot plasmas. Another possibility is that the hot flux rope was the heated rising cool filament since the latter was slightly lower than the former at the same time (compare panels (c) and (f) in \nfig{fig2}).

\begin{figure*}
\plotone{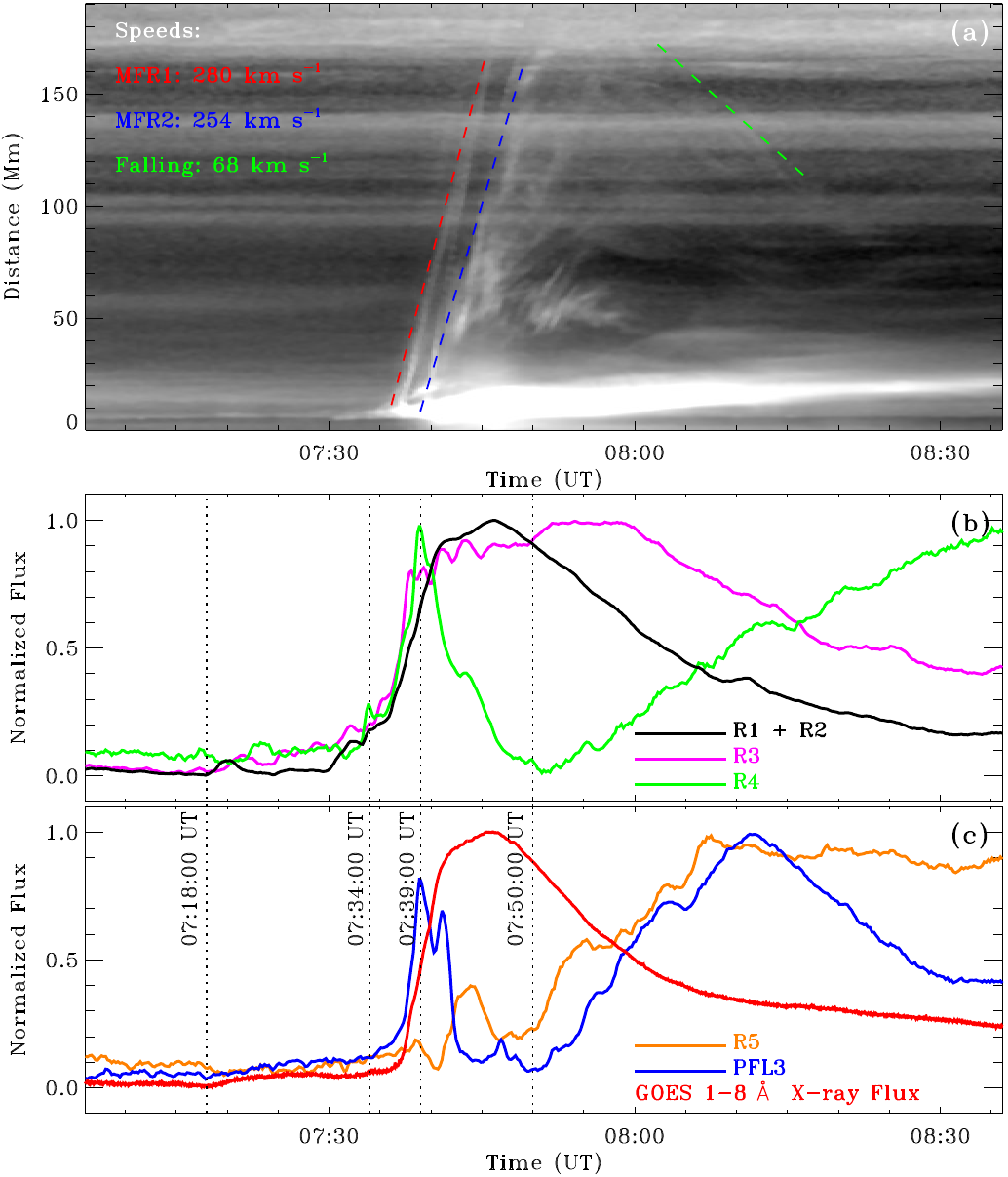}
\caption{Panel (a) is a time-distance diagram made along the eruption direction of the MFRs using the AIA 131 \AA\ time-sequence images, where the red and blue dashed lines are linear fit to the erupting flux ropes, and the green dashed line is the linear fit to the falling MFR. Panel (b) shows the lightcurves of R1+R2 (black), R3 (pink), and R4 (green), while panel (c) shows the lightcurves of the upper part (blue) and western footpoint (orange) of PFL3, respectively. The red curve in panel (c) is the {\em GOES} soft X-ray flux in the energy channel of 1 -- 8 \AA.~Four vertical dotted lines indicate some key times during the event.
\label{fig4}}
\end{figure*}

The formation and eruption details of the second MFR (hereafter called MFR2) are shown in \nfig{fig3}, which began at about 07:39 UT, 5 minutes after the start time of MFR1. In the H$\alpha$ images (the top row of \nfig{fig3}), four bright flare ribbons can also be observed at the same locations as those observed during the formation and eruption of MFR1, but they became more pronounced. Note that an additional bright flare ribbon (R5) appeared to the west of R4 (see the red arrow in \nfig{fig3} (c)). MFR2 appeared in the AIA 131 \AA\ image at about 07:39 UT (see the green arrow in \nfig{fig3} (d) and (g)). At this time, MFR1 can also be identified in the AIA 131 and 193 \AA\ images (see the red arrows in panels (d) and (g) in \nfig{fig3}). Such a picture reminds us of the double flux rope equilibrium configuration of the double-decker filament model proposed by \cite{2012ApJ...756...59L}. About 3 minutes later, MFR2 had risen by about 50 Mm (see the green arrow in \nfig{fig3} (e) and (h)), and the flare loop (hereafter called PFL2) had grown much larger relative to PFL1. It is noteworthy that when the erupting MFR2 started to fall back, another bright flare loop (hereafter called PFL3) appeared on the north side of PFL2. PFL3 appeared gradually, and its upper part reached the maximum brightness at about 08:10:54 UT, as shown in \nfig{fig3} (f). It should be a hot loop because it can not be observed at cooler wavelengths such as AIA 193 \AA\ images (see \nfig{fig3} (i) and the online animation). In the meantime, the western footpoint of PFL3 was also brightened, and it appeared as an elongated flare ribbon whose location is the same as the flare ribbon R5 observed in the H$\alpha$ image (see the red arrows in \nfig{fig3} (c), (f) and (i)). The eastern footpoint of PFL3 can not be recognized due to the contamination from PFL2. However, based on information on magnetic connectivity and arch characteristics of a coronal loop, we can speculate that the eastern footpoint of PFL3 should be close to the eastern footpoint of PFL2 because the magnetic flux at the east footpoint of PFL3 was negative. The appearance of PFL3 suggests that there was another magnetic reconnection occurring in the higher corona in addition to the magnetic reconnection that gave rise to PFL2. 

To investigate the detailed kinematics of the erupting MFRs, we made a time-distance diagram along the eruption direction by using the AIA 131 \AA\ images (see \nfig{fig4} (a)). The time-distance diagram clearly shows the successive eruptions of the two MFRs as two bright oblique stripes, and it can be seen that MFR1 and MFR2 started their fast-rising phase at about 07:34 UT and 07:39 UT, respectively. The rise of MFR2 reached a projection height of about 170 Mm and then began to fall back, but it is unclear whether the eruption of MFR1 was failed or successful because it became invisible at a higher altitude. By applying a linear fit to the bright stripes, it is obtained that the rising speeds of MFR1 and MFR2 are about \speed{280 and 254}, respectively. In addition, the falling speed is about \speed{68}. The intensity variations of the flare ribbons R1 -- R4 are shown in \nfig{fig4} (b), using intensity lightcurves measured from AIA 131 \AA\ images. In addition, the {\em GOES} soft X-ray flux in the 1 -- 8 \AA\ energy channel, the intensity variations of the upper part and the western footpoint (R5) of PFL3 are all plotted in \nfig{fig4} (c). It can be seen that the intensity of the flare ribbons R1 -- R4 started to increase at about 07:18 UT (12 minutes before the start of the {\em GOES} C4.6 flare). This time corresponds to the appearance of the small bright point near the intersection between F1 and F2 and the start time of the cancellation between the positive and negative photospheric magnetic fluxes (see \nfig{fig1} (h)). These results suggest that the formation of MFR1 might have started from the appearance of the small bright point and the cancellation of the opposite magnetic fluxes. The fast rise of MFR1 began at about 07:34 UT, delaying the onset of the {\em GOES} C4.6 flare by about 4 minutes. The fast rise of MFR2 began at about 07:39 UT, 5 minutes after the fast rise of MFR1. It is noted that the total intensity lightcurve of R1 and R2 peaked at about 07:45 UT, consistent with the peak time of the {\em GOES} C4.6 flare. The lightcurves of R3 and R4 also showed a similar variation trend as those of R1+R2 during the rising phase of the flare. However, the intensities of the upper part and the western footpoint (R5) of PFL3 showed a different variation pattern concerning the flare. They started to increase and reached their peaks at about 07:50 UT and 08:10 UT, respectively. Here, it should be pointed out that the intensity variations of the lightcurves of PFL3 and R5 before 07:50 UT were due to the contamination from the main flare emission, since before that, PFL3 and R5 had not started to form. These results are consistent with the imaging observations mentioned above. Therefore, the late appearance of PFL3 and R5 might imply another magnetic reconnection at a relatively higher altitude than the one producing the main flare. We further checked the LASCO CME Catalog and found a weak CME that erupted into the northeast quadrant of the outer corona and solar wind \footnote{\url{https://cdaw.gsfc.nasa.gov/movie/make_javamovie.php?stime=20220320_0857&etime=20220320_2211&img1=lasc2rdf&title=20220320.104805.p038g;V=113km/s}}, which may be due to the tether-cutting reconnection below MFR1 and therefore caused the appearance of PFL3 in the low corona.

\begin{figure*}
\plotone{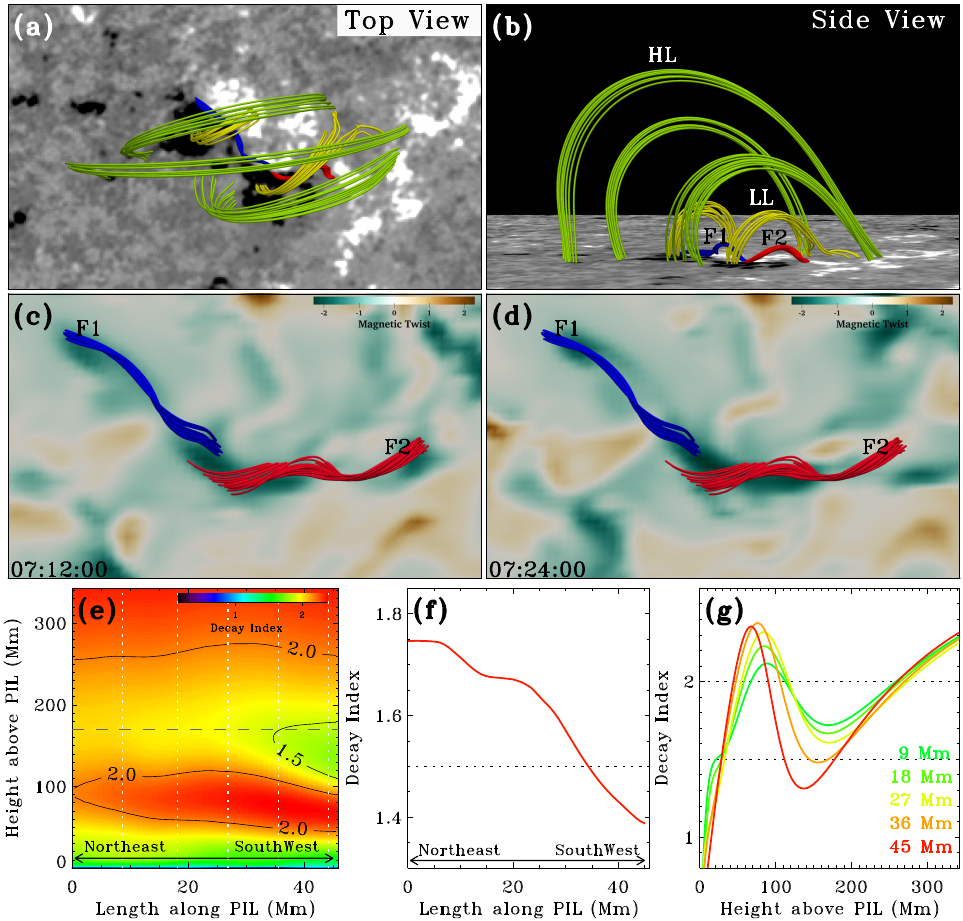}
\caption{Magnetic properties of the eruption source region. Panels (a) and (b) are top and side views of some extrapolated magnetic field lines. In each panel, the green and yellow curves are high-lying (HL) and low-lying (LL) loops, while the extrapolated F1 and F2 are shown in blue and red colors, respectively. Panels (c) and (d) show the calculated magnetic twist number maps at 07:12:00 UT and 07:24:00 UT, respectively. Panel (e) shows the magnetic decay index map in the vertical plane along the PIL, in which the black curves are the contours at a decay index of 1.5 and 2.0, and the magnitude of the decay index values at each point are with different colors. Panel (f) shows the decay index along the black horizontal dashed line in panel (e), while panel (g) shows the decay index profiles in height at five different locations, as indicated by the white vertical dashed lines in panel (e). 
\label{fig5}}
\end{figure*}

To better understand the formation and eruption of the newly formed MFRs, we further analyzed the three-dimensional coronal magnetic field over the eruption source region. The coronal magnetic field is extrapolated using the nonlinear force-free magnetic field (NLFFF) optimization method, which assumes a static configuration without Lorentz force, i.e., the current density is parallel to the magnetic field everywhere \citep{2000ApJ...540.1150W}.  In this paper, the calculation was performed in a Cartesian volume of $832 \times 792 \times 792$ pixels, corresponding to a physical volume of about $361 \times 343 \times 343$ Mm$^3$. We extrapolated the coronal magnetic field at 07:12:00 UT before the eruption by using the photospheric vector magnetic field as the bottom boundary. Since the input photospheric vector magnetic map used as the bottom boundary condition is not force-free, we first preprocessed it to best fit the force-free condition according to the method given in \cite{2006SoPh..233..215W}. To quantify the quality of the preprocessed vector magnetogram, we further calculated the three dimensionless parameters introduced in \cite{2006SoPh..233..215W}. The calculation results show that the values of the flux balance, force balance, and torque balance dimensionless parameters are about $-5.6 \times 10^{-2}$, $2.9 \times 10^{-3}$, and $7.1 \times 10^{-2}$, respectively. These results are all far less than 1, so they satisfy the force-free assumption of the NLFFF extrapolation method \citep{2006SoPh..233..215W}. For an ideal force-free magnetic field, the values of these parameters should be 0. In practice, however, non-zero finite values of these parameters are acceptable due to the influences resulting from the numerical calculation and the real input magnetogram data. In general, a good NLFFF extrapolation requires that the values of the volume-averaged fractional flux and the current-weighted angle between the modeled magnetic field and electric current density are far less than $20 \times 10^{-4}$ and $10^{\circ}$, respectively \citep[e.g.,][]{2015ApJ...811..107D, 2019ApJ...887...64T}. Based on the extrapolated three-dimensional magnetic field, we calculated the volume-averaged fractional flux \citep{2000ApJ...540.1150W, 2015ApJ...811..107D} and the current-weighted angle between the modeled magnetic field and electric current density \citep{2006SoPh..235..161S} to quantify the quality of the extrapolated magnetic field \citep[e.g.,][]{2015ApJ...811..107D, 2019ApJ...887...64T, 2020ApJ...900...23M, 2021MNRAS.501.4703J}. It is found that the values for the volume-averaged fractional flux and the current-weighted angle between the modeled magnetic field and electric current density are of about $3.3 \times 10^{-4}$ and $9.7^{\circ}$, respectively. These results indicate that the extrapolated magnetic field meets the criteria of a good NLFFF extrapolation.

The analysis results based on the extrapolated magnetic field are shown in \nfig{fig5}, where panels (a) and (b) show the top and side views of the magnetic connection in the eruption source region, respectively. It can be seen that the extrapolated three-dimensional coronal magnetic field well reveals some key magnetic structures, including F1 (blue), F2 (red), and their corresponding confining magnetic field lines (yellow and green curves). These extrapolated coronal magnetic structures are in good agreement with the observational results revealed by the H$\alpha$ and the AIA EUV images. We measured the heights of the filaments and the confining loops based on the extrapolated magnetic field. The results show that the heights of F1 and F2 are both about 4.5 Mm from the photosphere, while the heights of the low-lying and high-lying loops, as shown in \nfig{fig5} (b), are about 13.5 and 50 Mm from the photosphere, respectively. It should be noted that only a few representative field lines are shown in the figure.

We also calculate the magnetic twist number maps at a chromospheric height of about 1.5 Mm from the photosphere based on the extrapolated three-dimensional coronal magnetic fields at 07:12:00 UT (before the flare start) and 07:24:00 UT (after the flare start). The calculation is based on the code developed by \cite{2016ApJ...818..148L}, where the authors used the equation $T_{w} = \int_{L} \frac{\nabla~\times~{\bf B}~\cdot ~{\bf B}}{4\pi B^{2}}~dl$ = $\frac{1}{4\pi}\int_{L}~\alpha~dl$, if $\nabla~\times~{\bf B} = \alpha {\bf B}$ \citep{2006JPhA...39.8321B}. Here, $\alpha$ is the force-free parameter, and ${\nabla~\times~{\bf B}~\cdot ~{\bf B}}/{4\pi B^{2}}$ can be regarded as the local twist density along the magnetic field line. The calculated results are shown in \nfig{fig5} (c) and (d), where the extrapolated F1 and F2 are also plotted as blue and red colors, respectively. The average twist number along F1 and F2 are calculated using the method proposed in \cite{2016ApJ...818..148L}. The calculation results show that the average twist number of F1 (F2) at 07:12 UT and 07:24 UT are about -1.26 (1.73) and -1.68 (1.83) for F1 (F2), respectively. These results indicate that the photospheric shear and convergence motions of the opposite magnetic polarities provided additional magnetic twists to the filaments, and the magnetic twist numbers of the filaments at 07:24:00 UT are smaller than the minimum average twist number for the kink instability \citep[2 turns,][]{2019ApJ...884...73D}. These results may suggest that the initiation of F1 and F2 was probably mainly due to the photospheric motions around their inner footpoints, and the magnetic reconnection between the two filaments can naturally transfer magnetic twists into the newly formed MFRs. 

To figure out the influence of the background magnetic field on the MFR eruptions, we checked the decay index of the coronal magnetic field above the PIL. In general, the decay index characterizes the rate of decrease of the transverse component of the potential field with height, and its mathematical expression can be written as $n = -d~{\rm ln}~B_{t}/d~{\rm ln}~ h$  \citep{2006PhRvL..96y5002K}. Here, $h$ and $B_{t}$ are the height above the photosphere and the transverse component of the potential field, respectively. In principle, a higher decay index value of the overlying coronal magnetic field implies a greater possibility of an MFR suffering from torus instability. Therefore, a higher decay index implies a greater possibility of an MFR to erupt successfully as a CME.  \cite{2006PhRvL..96y5002K} theoretically pointed out that an MFR becomes torus unstable when the decay index of the external poloidal field is in the range of 1.5 -- 2.0. However, in practice, a few observations also found that the eruption of some torus-unstable MFRs also failed to cause CMEs, especially when the erupting MFRs show strong rotation motions with large rotation angles ranging from 50$^\circ$ to 130$^\circ$ \citep{2019ApJ...877L..28Z}. To calculate the decay index in a real observation, the transverse component of the extrapolated potential field is often used to approximate the external poloidal field at the axis of the MFR \citep[e.g.,][]{2007AN....328..743T, 2011RAA....11..594S, 2013ApJ...779..129K}. In the present study, following the analysis method used in \cite{2021SoPh..296...85J} and \cite{2022ApJ...926..143M}, we calculate the decay indexes in the vertical plane along the PIL of the active region AR12971, and the results are plotted in \nfig{fig5} (e)--(g). In the decay index map (\nfig{fig5} (e)), the black curves are contours at the values of 1.5 and 2.0, and different colors indicate different decay index values. It can be seen that the distribution of the decay index of the magnetic field above the PIL could be divided into four distinct regions of different decay index values, as indicated by the contour lines at the value of 2.0. For the low decay index region in the height range of about 100 -- 250 Mm from the photosphere, it can be seen that the decay index becomes progressively lower from the northeast to the southwest. We further plot the decay index profile along the horizontal black dashed line (see \nfig{fig5} (e)) in \nfig{fig5} (f), which shows the decreasing trend of the decay index more clearly. This particular decay index distribution above the PIL may explain the oblique eruption of the MFRs towards the northeast direction since a higher decay index implies a greater possibility for an eruption flux rope to suffer from torus instability.

To analyze the details of the decay index distribution in the vertical direction, we further examine the decay index profiles at five specific locations along the PIL as indicated by the white vertical dashed lines in \nfig{fig5} (e). One can see in \nfig{fig5} (g) that the distribution of the decay index profiles shows a so-called saddle-like profile where a low decay index region is sandwiched between two high decay index regions \citep[e.g.,][]{2017ApJ...843L...9W, 2022ApJ...926..143M}. For the present study, we note that from the bottom up, the increasing trend of the decay index changes to a decreasing trend at a height of about 80 Mm from the photosphere (see the decay index profile at the distance of 27 Mm from the northeast end of the PIL in \nfig{fig5} (g)). This decreasing trend stops at a height of about 170 Mm from the photosphere, and then the decay index starts to increase monotonically with height. Since the heights of the filaments and their confining loops are all below the height of 50 Mm, the existence of the low decay index region in the height range of 100 -- 250 might be important for the failed eruption of the MFRs, since a lower decay index implies a reduced possibility for a flux rope to erupt successfully. As can be seen in the time-distance diagram (see \nfig{fig4} (a)), the erupting MFR2 started to fall back at a projection height of about 170 Mm. Since the MFRs erupted obliquely towards the northeast direction, we can assume that the eruption angle of inclination was about $45^\circ$ concerning the vertical direction. Therefore, the measured projection distance of the erupting MFRs equal to its vertical height from the photosphere; they are both 170 Mm. This result supports the idea that the erupting MFRs experienced a strong constraint at the height of around 170 Mm from the photosphere because this height corresponds to the bottom of the saddle of the decay index profile. In previous studies, some authors found that the existence of a saddle-like decay index profile of the coronal magnetic field above the eruption source region creates a favorable condition for the generation of a failed flux rope eruption since this means that the toroidal strapping force increases in height after an initial decrease \citep{2017ApJ...843L...9W, 2020ARep...64..272F, 2022ApJ...926..143M}. However, as a flux rope eruption is often affected by a variety of physical factors, a saddle-like decay index profile magnetic system could also produce successful eruptions in practice \citep[e.g.,][]{2017ApJ...843L...9W, 2018NatCo...9..174I, 2020MNRAS.494.2166F}.

\begin{figure*}
\plotone{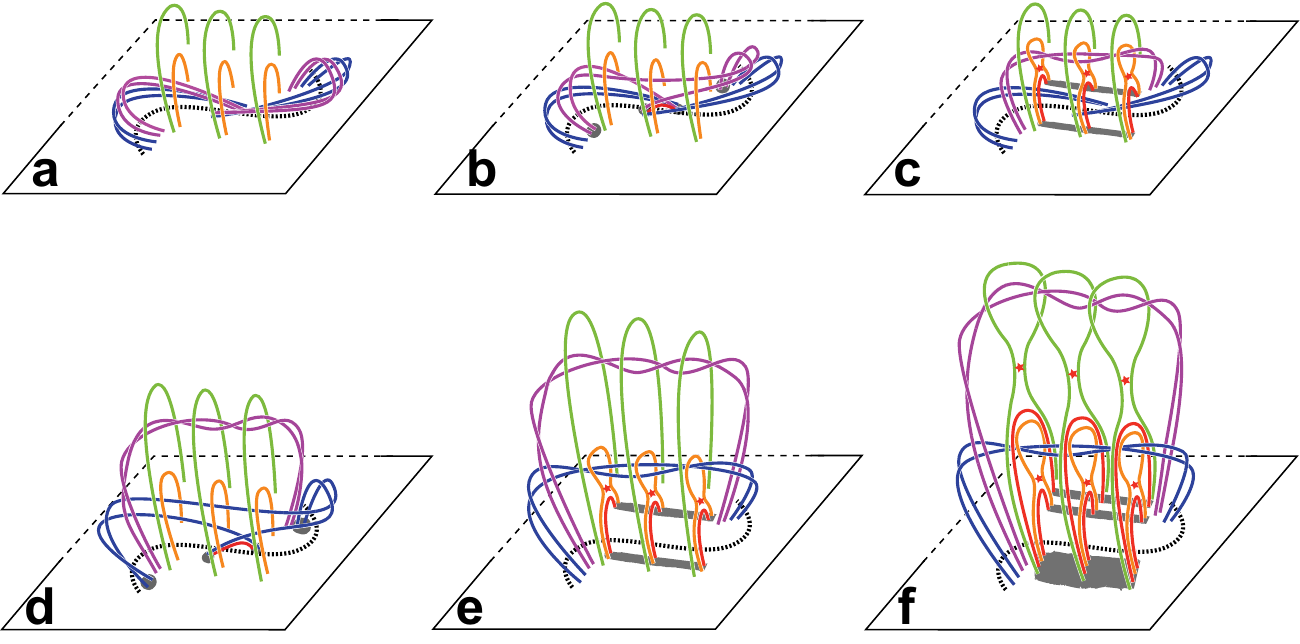}
\caption{A cartoon model illustrates the formation mechanism of the double-decker filament system (a double flux rope type). The figure shows only a few representative field lines. In each panel, the dashed black curve is the PIL in between the opposite magnetic polarities, and the yellow and green curves are the low-lying and high-lying confining loops of the filaments (represented by the two sets of J-shaped curves), respectively. The twisted fuchsia and indigo curves in panels c -- f represent the newly formed MFR1 and MFR2, respectively. The red stars indicate the magnetic reconnection positions, the red curves represent the newly formed PFLs, and the gray features mark the flare ribbon locations.
\label{fig6}}
\end{figure*}

\section{Interpretation and Discussion}
Based on the observational results, it is clear that the present event can be explained within the framework of the so-called tether-cutting model that was first proposed by \cite{1980IAUS...91..207M} and then confirmed by more detailed observations \citep{2001ApJ...552..833M} and sophisticated numerical simulations \citep[e.g.,][]{2021NatAs...5.1126J}. The tether-cutting model describes both confined (failed) and ejective (successful) solar eruptions originating in simple bipolar regions where the core field is highly sheared and twisted. In general, for a simple bipolar magnetic region, the positive and negative polarities are often connected by a potential field to maintain the lowest energy state. However, due to the photospheric activities such as the shear motion parallel to the PIL and the convergence motion perpendicular to the PIL, the bipolar magnetic system becomes more and more non-potential so that it is prone to erupt for releasing magnetic free energy. In observation, the core field of the bipole shows as a sigmoidal structure \citep[composed of two opposite J-like bundles of loops and shaped like a forward S or inversed S;][]{1996ApJ...464L.199R, 2008A&A...481L..65M} often seen in soft X-ray and sometimes in EUV images \citep[e.g.,][]{2001ApJ...552..833M, 2010ApJ...725L..84L}. In chromospheric observations such as the H$\alpha$ line, the shear core often appears as two overlapping inverse-S-shaped filaments along the photospheric PIL and approximately aligned to the central portion of the sigmoid observed in soft X-ray and EUV emission \citep{2002SoPh..207..111P}. Many observations have shown that sigmoidal regions are more prone to erupt than non-sigmoidal regions without significant dependence on size scales \citep[e.g.,][]{2007ApJ...671L..81C, 2000GeoRL..27.2161G, 2010ApJ...718..981R, 2013MNRAS.431.1359Z}. 

According to the tether-cutting model, the arms of the two opposite J-shaped loops shear past each other along the middle portion of the PIL. They are poised to reconnect and therefore lead to the instability and even eruption of the sigmoidal system when and if they come into contact and push against each other. The reconnection between the sheared opposite J-shaped loops produces simultaneously downward- and upward-moving hot reconnected magnetic field lines. Meanwhile, the particle beams accelerated in the magnetic reconnection move downward along the newly formed reconnected magnetic field lines from the middle portion and collide with the dense plasma at the footpoints of the loops to form four bright emission patches. It should be pointed out that the two inner conjugated bright patches connected by the short, downward-moving reconnected loops often appear as a compact bright patch in low spatial resolution observations because they are too close together to be distinguished. The long upward-moving reconnected loops often show as a twisted MFR that connects the far ends of the two elbows of the two opposite J-shaped structures. As the newly formed MFR rises continuously, magnetic field lines rooted farther away from the shearing core (acting as the confining field of the rising MFR) are stretched outward. This will lead to the formation of a vertical current sheet below the rising MFR, and magnetic reconnection within it eventually causes the system to erupt, leaving behind two elongated flare ribbons connected by a bright PFL system striding the PIL, as depicted in standard flare models \citep[e.g.,][]{2000JGR...105.2375L, 2012A&A...543A.110A}. In addition, the eruption of the newly formed MFR can also be arrested and confined within the closed bipolar system, i.e., failed to open the closed bipolar field, although it starts the same way as the successful eruption of a sigmoidal structure.

Based on our analysis results, we interpret the present event using the tether-cutting model and emphasize that a double-decker flux rope system can be formed through two successive confined tether-cutting eruptions in the same bipolar active region. To explain the detailed evolution process of the eruption, we draw a cartoon in \nfig{fig6}, in which only some representative magnetic field lines are shown. \nfig{fig6} (a) shows the pre-eruption magnetic configuration of the eruption source region, in which the black dashed curve represents the PIL in-between the two opposite magnetic polarities, while the fuchsia and indigo curves composed of the two overlapping inverse-S-shaped filaments (or loops in EUV and X-ray observations) confined by the overlying yellow and green potential loops that stride the PIL. Due to the combined effect of the photospheric shearing motion parallel to the PIL and the convergence motion perpendicular to the PIL around the middle portion of the PIL, the arm part of some magnetic field lines composing the two overlapping inverse-S-shaped filaments come into contact and push against each other. Therefore, an X-type current sheet forms between the arms of the two overlapping inverse-S-shaped filaments, and magnetic reconnection in this current sheet will lead to two groups of newly formed reconnected loops, i.e., the downward-moving short flare loops striding the PIL  (see the red loop in \nfig{fig6} (b)) and the upward-moving long MFR connecting the far ends of the two overlapping inverse-S-shaped filaments along the PIL (see the long fuchsia loop in \nfig{fig6} (b)). In addition, the reconnection also produces four flare ribbons at the footpoints of the downward-moving flare loops and the upward-moving MFR (see the gray patches in \nfig{fig6} (b)), as the particle beams accelerated by the magnetic reconnection stream downward along the reconnected field lines and collide with the dense plasma at the footpoints of the loops. During this phase, the magnetic twists stored in the two overlapping inverse-S-shaped filaments are transferred to the newly formed MFR. As the long reconnected loops continuously rise and form, they together form a twisted MFR that stretches the low-lying confining magnetic field lines in the vertical direction and leads to the formation of a vertical current sheet between the two legs of the confining loops (see the fuchsia twisted loops in \nfig{fig6} (c)). The magnetic reconnection in this newly current sheet removes some of the confining fields of the rising MFR, leading to its further rise. However, the rise of the MFR is not sustainable due to the existence of the low magnetic decay index region above the PIL. It will be slowed down or even stopped at a certain height around the low decay index region as shown in \nfig{fig5}(e), and eventually reaches a new equilibrium at a relatively higher height in the low corona. In addition, the magnetic reconnection also leads to the formation of a group of flare loops striding the PIL and two conjugated flare ribbons on either side of the PIL (see the red loops and the two gray rectangles in \nfig{fig6} (c)). 

A few minutes later, probably due to the resumption of the rapid increase of the negative magnetic flux around the middle portion of the PIL as evidenced in \nfig{fig1}(h), the magnetic system might gather enough energy and appropriate magnetic condition to erupt again. Therefore, the system starts the formation and eruption of MFR2 with the same evolution process as MFR1 (see \nfig{fig6} (d) and (e)). Around the moment after the formation of MFR2 (see \nfig{fig6}(c)), both MFR1 and MFR2 are confined within the bipolar magnetic system but separated from each other in height (also see \nfig{fig3}(d) and (g)), resembling the physical picture of a double-decker filament system as defined by \cite{2012ApJ...756...59L}. This suggests that two MFRs formed successively from the same PIL through confined tether-cutting eruptions can form a double-decker filament system (double flux rope type). Note that in the present event, there is a 5-minute pause between the start times of the fast rise of the two MFRs from 07:34 UT to 07:39 UT. This time interval is consistent with the shortstop or slowdown period of the fast increase of the negative magnetic flux around the middle portion of the PIL (see \nfig{fig1} (h)). In particular, the onset time of MFR2 was the same as the restart time of the fast increase of the negative magnetic flux, which probably demonstrates that the formation of the observed MFRs was closely related to the photospheric activities that drive the fast increase of the negative magnetic flux and the flux cancellation around the middle portion of the PIL. 

In comparison with our observations, we observed all the expected characteristic signals predicted by the tether-cutting model, including a small brightening at the intersection of the arms of the two overlapping inverse-S-shaped filaments or loops before the main flare, flare ribbons at the footpoints of the PFL, and the newly formed MFR. It should be noted that during the magnetic reconnection within the X-type current sheet between the arms of the two overlapping inverse-S-shaped filaments, the conjugated two flare ribbons connected by the flare loop are difficult to distinguish. They are often observed together as a small compact bright point in H$\alpha$ and EUV images because they are too small and so close to each other (see the red arrows in \nfig{fig2} (a) and (d)). In addition, the magnetic reconnection at the intersection of the two overlapping inverse-S-shaped filaments can last for a relatively long time before the onset of the main flare, and the resulting brightening can be considered to be an important precursor for solar eruptions. For example, in the present event, the small brightening started about 36 minutes before the main flare. During this phase, the predicted upward-moving MFR should also start to form, which is not observable except for some disturbance or activation signals in the filaments. The flare loop and the two associated flare ribbons during the magnetic reconnection in the vertical current sheet are more pronounced and can be observed. For example, PFL1 and PFL2 shown in \nfig{fig2} (f) and \nfig{fig3} (f) are associated with the magnetic reconnection in the vertical current sheets below MFR1 and MFR2, respectively. As shown in \nfig{fig3} (f) and \nfig{fig4} (c), there was a group of additional higher flare loops (PFL3) and a flare ribbon (R5) on the western edge of the bipole. We interpret these features as the production of the magnetic reconnection between the two legs of the higher confining loops. After MFR1 escaped the confinement of the lower confining magnetic field, it continued to rise but was also confined by the higher confining loops (see the green loops in \nfig{fig6} (c) to (f)). When MFR1 reaches a certain height, these higher loops can also collapse in the middle to form a vertical current sheet, thus triggering an additional magnetic reconnection between MFR1 and MFR2 (see the red asterisks in \nfig{fig6} (f)). This also implies that MFR1 might be a successful eruption that became the core of the observed weak CME in the northeast quadrant of the outer corona. However, if this is the truth, MFR1 should also experience a short pause before its successful eruption, due to the existence of the saddle-like distribution of the magnetic decay index in the vertical direction. That is to say, at least for a period, the pair of newly formed MFRs can be regarded as a double-decker flux system. The observed PFL3 should correspond to the red curves located between the green and yellow curves, while the observed flare ribbon (R5) should correspond to the outermost ribbon as shown by the gray rectangle feature in \nfig{fig6} (f). In practice, this reconnection should produce two conjugated flare ribbons at both footpoints of PFL3. However, we only observed the western one in observations. The other one should be merged with the eastern flare ribbon associated with PFL2, as shown by the thick gray rectangle feature in \nfig{fig6} (f). 

What kind of physical factors determine the outcome of a solar eruption has attracted much attention from solar physicists. In principle, the failure or success of a solar eruption depends on the outcome of the competition between upward and downward forces. If the upward force, such as magnetic pressure, is smaller than the downward forces, such as gravity and magnetic tension, the eruption is often a failed eruption. In recent years, many possible physical factors have been proposed to explain failed filament eruptions \citep[e.g.,][]{2003ApJ...595L.135J, 2011RAA....11..594S, 2013ApJ...778...70C, 2018ApJ...858..121L,2021Univ....7..405Z, 2022SoPh..297...81J, 2022ApJ...941L...1L, 2022ApJ...929L..23H, 2022A&A...665A..51S}, especially those that take into account the physical properties of the overlying magnetic field, for example, a low magnetic field strength at low altitudes \citep[e.g.,][]{2008ApJ...679L.151L, 2009ApJ...696L..70L, 2018Natur.554..211A,2022A&A...665A..51S}, a small gradient of the overlying magnetic field strength with respect to the height \citep[e.g.,][]{2006PhRvL..96y5002K, 2015ApJ...814..126Z, 2012ApJ...750...12S}. Some authors emphasize physical factors below the confining field, such as the influence due to insufficient kinetic energy \citep{2005ApJ...630L..97T, 2011RAA....11..594S, 2018ApJ...858..121L} and the rotation angle of the erupting filaments \citep[e.g.,][]{2018ApJ...853..105B, 2019ApJ...877L..28Z}. These possible physical factors may act separately or in combination to cause a failed filament eruption, and filament eruptions within different magnetic configurations may be dominated by different influencing factors. For example, for a failed filament eruption in a tether-cutting topology, a weak magnetic shear core relative to the overlying magnetic field and a higher reconnection height between the two opposite J-shaped loops could be important for a failed filament eruption in a tether-cutting topology \citep{2001ApJ...552..833M}. In addition, eruptions in a magnetic breakout topology \citep[e.g.,][]{1999ApJ...510..485A,2012ApJ...750...12S} can also fail if the external magnetic reconnection occurs between the two groups of lateral closed loops because the two lateral low-lying closed loops continuously transform into the envelope confining loops and thus provide additional magnetic confinement to the erupting core field \citep{2022A&A...665A..51S}. 

For the present event, the failed eruption of the newly formed MFR2 and the possible short pausing of MFR1 could be caused by several different physical factors. In principle, as mentioned above, the magnetic reconnection between the two legs of the confining field lines of the rising MFR will remove some of the confining field lines, thus reducing the magnetic confinement of the MFR. In a normal coronal magnetic field with a decay index increasing monotonically with height, this will cause positive feedback to the rising MFR and eventually lead to a successful eruption. However, in a coronal magnetic field whose decay index has a saddle-like profile in height, as in the present event, the rising MFR is more strongly confined by the overlying coronal magnetic field around the height of the saddle base \citep{2022ApJ...929....2L}. Naturally, the erupting MFR could be slowed down and/or stopped there, eventually reaching a new equilibrium in the high corona. Therefore, the saddle-like distribution of the magnetic decay index in height could be an important physical factor for the failed eruption of the MFR2, as well as the pausing of MFR1 before its successful eruption. In the practical observations, although the final equilibrium state of the double-decker flux ropes system in the present case is not detected to keep stable for a long time in the H$\alpha$ and EUV observations, some previous studies do show the fact that erupting filaments or flux ropes can indeed maintain a new equilibrium around the height of the saddle base for a long time \citep[e.g.,][]{2016ApJ...821...85G, 2017SoPh..292...81C, 2018MNRAS.475.1646F}. This suggests that the formation of a double-decker filament system through two successive confined tether-cutting eruptions from the same PIL is possible if the background coronal magnetic field has a saddle-like distribution of the magnetic decay index in height.

As suggested by \cite{2006PhRvL..96y5002K}, an MFR becomes torus unstable when the decay index of the external poloidal field is in the range of 1.5 -- 2.0. In our case, the decay index within the saddle region is generally less than 2.0 but greater than 1.5, except for the small low decay index region above the southwest part of the PIL (see \nfig{fig5} (e) and (f)). This raises the question of why a torus unstable decay index region prevents the successful eruption of the MFRs in the present case. We argue that the combination of multiple physical factors probably caused the pause and failed eruptions of the newly formed MFRs. Besides the constraint influence of the low decay index region above the PIL, there are at least two other physical factors that might contribute to the pause and failed eruptions of the MFRs. First, as suggested in \citep{2011RAA....11..594S}, the energy obtained by the erupting MFRs may be insufficient to overcome the confinement of the overlying magnetic field because the accompanying flare of the present event was only a small energyless {\em GOES} C4.6 flare. The second reason may be due to the weak magnetic core field, which was composed of two small overlapping inverse-S-shaped filaments. Because for confined tether-cutting eruptions, a weak magnetic shear core relative to the overlying magnetic field should have less possibility to erupt successfully \citep{2001ApJ...552..833M}. In addition to these possible physical factors, the erupting MFR2 is still affected by the increased magnetic confinement of its overlying confining magnetic field due to the continuous generation of PFL3 by magnetic reconnection below MFR1 but above MFR2 (see the red loop in between the green and the yellow loops in \nfig{fig6} (f)).

The magnetic reconnection in-between MFR1 and MFR2 is similar to the internal reconnection inside an MFR containing a filament at the bottom, which often causes the successful (failed)  eruption of the upper (lower) part as seen in many partial filament flux rope eruptions \citep[e.g.,][]{2001ApJ...549.1221G, 2012ApJ...750...12S}. In previous observations, many authors have proposed that a double-decker filament could be formed by splitting a pre-existing single filament \citep[e.g.,][]{2012ApJ...756...59L,  2018NewA...65....7T, 2022ApJ...929...85D, 2022ApJ...933..200Z}. However, for the present case, although we observed the reconnection between the two MFRs as evidenced by the appearance of PFL3, our observational results indicate that the formation of the double-decker pair of flux rope system was not due to the splitting of a single filament or flux rope. We propose that the transient double-decker pair of flux rope system was formed by two successive MFR eruptions formed in the same PIL. Furthermore, in published cases of double-decker filaments formed by internal reconnection, the upper branch is often observed as a successful eruption, while the state of the lower branch could be either stable or erupted as a failed or successful eruption \citep[e.g.,][]{2012ApJ...750...12S, 2014ApJ...789...93C, 2021ApJ...923..142C, 2018ApJ...860...35D}. Partially, in some cases, the eruption of the lower branch may overtake the upper one \citep{2018NewA...65....7T}, or the rise of the lower branch is accompanied by the descent of the upper one \citep{2015ApJ...813...60Z}. In these cases, the two branches are observed to interact and merge into one single filament, where they erupt together to cause a single CME in the outer corona. In our case, MFR2 should be a failed eruption, but MFR1 may erupt successfully after a short pausing at the height of the saddle base of the decay index in height.

It should be noted that many studies only consider the decay index in the vertical or radial direction. This can be inaccurate, as many flux rope eruptions are often not along the vertical or radial direction, such as in the present case. As shown by the decay index map in \nfig{fig5} (e), the erupting MFRs tend to avoid the low decay index region and erupt towards the relatively high decay index region. Therefore, the spatial decay index distribution of the coronal magnetic field above the PIL may have a significant influence on the eruption direction of MFRs.

\section{Summary}
Using high spatiotemporal resolution and multi-wavelength observations taken by the NVST and the {\em SDO}, we report the successive formation and subsequent eruption of two MFRs along the same PIL of AR12971 on 2022 March 20. We propose that a double-decker flux rope system, consisting of a double flux rope equilibrium, can be formed by two successive confined tether-cutting eruptions from the same PIL. 

Our observations show that the formations of two MFRs are both due to the magnetic reconnection between the two overlapping inverse-S-shaped filaments along the same PIL, and the combined effect of the photospheric shearing motion parallel to the PIL and the convergence motion perpendicular to the PIL may play an important role in triggering of the magnetic reconnection by pushing the arms of the two overlapping inverse-S-shaped filaments against each other. This result is supported by the four observed flare ribbons at the footpoints of the newly formed MFRs and the PFLs striding the middle portion of the PIL. After the formation of each MFR, it erupted obliquely in the northeast direction, leaving behind a pair of conjugated flare ribbons connected by a group of bright PFL striding the PIL. These observational results can be interpreted by using the tether-cutting model \citep[confined case,][]{2001ApJ...552..833M}. More importantly, the observations and interpretations present in this paper also provide a new physical mechanism for understanding the formation of a double-decker filament system (the double flux rope type). According to this line of thinking, one can expect the formation of a triple-decker or multiple-decker filament or flux rope system if three or multiple MFRs intermittently form and erupt from the same PIL.

We find that the magnetic decay index of the coronal magnetic field above the PIL has a saddle-like profile in height. In contrast, along the PIL within the saddle, the decay index shows a gradual decreasing trend from the northeast end to the southwest end of the PIL. We propose that the particular distribution of the decay index of the coronal magnetic field above the PIL affects the erupting MFRs in two ways, i.e., the saddle-like decay index distribution in height can contribute to the failed eruption of the MFRs, while the asymmetric decay index distribution within the saddle along the PIL can lead to the non-radial eruption of the MFRs because an MFR always tends to erupt towards regions with smaller magnetic confinement. The failed eruption of  MFR2 and the possible pausing of MFR1 in the present event were probably attributed to the combined effect of several different possible reasons, including the saddle-like decay index distribution of the overlying coronal magnetic field in height, the small energyless accompanying flare, and the relatively weak core field of the magnetic system in the eruption source region. For a full understanding of the formation of double-decker filaments, further observational and theoretical studies are desirable in the future.

\vspace{0.4cm}
The authors thank the excellent data provided by the NVST and the {\em SDO} teams and the anonymous referee for his/her valuable suggestions and comments. This work was supported by the Natural Science Foundation of China (12173083, 11922307, 11773068), the Yunnan Science Foundation for Distinguished Young Scholars (202101AV070004), the National Key R\&D Program of China (2019YFA0405000), and the Specialized Research Fund for State Key Laboratories. Ahmed Ahmed Ibrahim would thank the Researchers Supporting Project number (RSPD2023R993), King Saud University, Riyadh, Saudi Arabia.


\end{document}